\documentclass[utf8]{FrontiersinHarvard} 

\usepackage{url,hyperref,lineno,microtype,subcaption}
\usepackage[onehalfspacing]{setspace}
\usepackage{braket}
\usepackage{booktabs}
\usepackage{graphicx}

\def\keyFont{\fontsize{8}{11}\helveticabold }
\def\firstAuthorLast{Chen {et~al.}} 
\def\Authors{Mengyu Chen\,$^{1,*}$, Marko Peljhan\,$^{1}$ and Misha Sra\,$^{1,2}$}

\def\Address{$^{1}$Department of Media Arts and Technology, University of California Santa Barbara, Santa Barbara , CA, United States \\
$^{2}$Department of Computer Science, University of California Santa Barbara, Santa Barbara, CA, United States}
\def\corrAuthor{Corresponding Author}

\def\corrEmail{mengyuchen@ucsb.edu}

\begin{document}
\onecolumn
\firstpage{1}

\title[EntangleVR++]{EntangleVR++: Evaluating the Potential of using Entanglement in an Interactive VR Scene Creation System}

\author[\firstAuthorLast ]{\Authors} 
\address{} 
\correspondance{} 

\extraAuth{}
\newcommand\added[1]{\textcolor{black}{#1}}

\maketitle

\begin{abstract}

\section{}
Interactive digital stories provide a sense of flexibility and freedom to players by allowing them to make choices at key junctions. These choices advance the narrative and determine, to some degree, how the story evolves for that player. As shown in prior work, the ability to control or participate in the construction of the narrative can give the player a high level of agency that results in a stronger sense of immersion in the narrative experience. To support the design of this type of interactive storytelling, our system, EntangleVR++, borrows the idea of entanglement from quantum computing. Our use of entanglement allows creators and storytellers control over which sequences of story events take place in correlation with each other, initiated by the choices a player makes. In this work, we evaluated how well our idea of entanglement enables creators to easily and quickly design interactive VR narratives. We asked 16 participants to use our system and based on user interviews, analyses of screen recordings, and questionnaire feedback, we extracted four themes. From these themes and the study overall, we derived four authoring strategies for tool designers interested in the design of future visual interface for interactively creating virtual scenes that include relational objects and multiple outcomes driven by player interactions. 

\tiny
 \keyFont{ \section{Keywords:} virtual reality, quantum computing, entanglement, visual programming, art, creativity, interactive narratives, scene creation} 
\end{abstract}

\section{Introduction}

Virtual reality (VR) has seen significant growth in recent years and demand for new types of experiences is rising. VR can allow players to experience the impossible such as step into Van Gogh's vividly colored paintings \footnote{https://www.exhibitionhub.com/exhibitions/van-gogh-immersive-experience/}, take an intimate look into an immigrant family's struggles during the war \footnote{https://www.randallokita.com/the-book-of-distance}, or explore the insides of the human body \footnote{https://www.oculus.com/experiences/rift/967071646715932/}. Many VR artworks such as Carne y arena \footnote{https://www.lacma.org/art/exhibition/alejandro-g-inarritu-carne-y-arena-virtually-present-physically-invisible} and La Camera in Sabiatta \footnote{https://www.e-flux.com/announcements/165304/laurie-anderson-hsin-chien-huangla-camera-insabbiata/} and short-film experiences such as Wilde Eastern \footnote{http://wildeeasternvr.com/} and Project Syria \footnote{https://docubase.mit.edu/project/project-syria/} have begun to use this new medium to present new types of interactive narratives. These experiences are often driven by the player which allows them to explore the immersive environments via simple interactions such as touching and picking up virtual objects, or making eye contact with virtual characters. However, in most of these experiences, the player's interactions do not have any impact on the virtual landscape or the story-line, as the main narrative components tend to be pre-scripted by the authors, with often a linear progression. Even in highly interactive VR games, the narratives are usually pre-scripted with no lasting cause-effect relationships leading to a similar experience for all players with a ``one size fit all'' approach. We believe there is potential for first person VR experiences, much like real life, to be individualized with dynamic and unpredictable paths and different endings based on a player's interactions with virtual world elements. The ability to construct unique player-centered experiences in VR with simple mechanisms that support a multiplicity of outcomes that depend on player choice, can help reinvent the text-based design space of \textit{Choose Your Own Adventure} narratives in immersive 3D. 

\begin{figure}
\includegraphics[width=\textwidth]{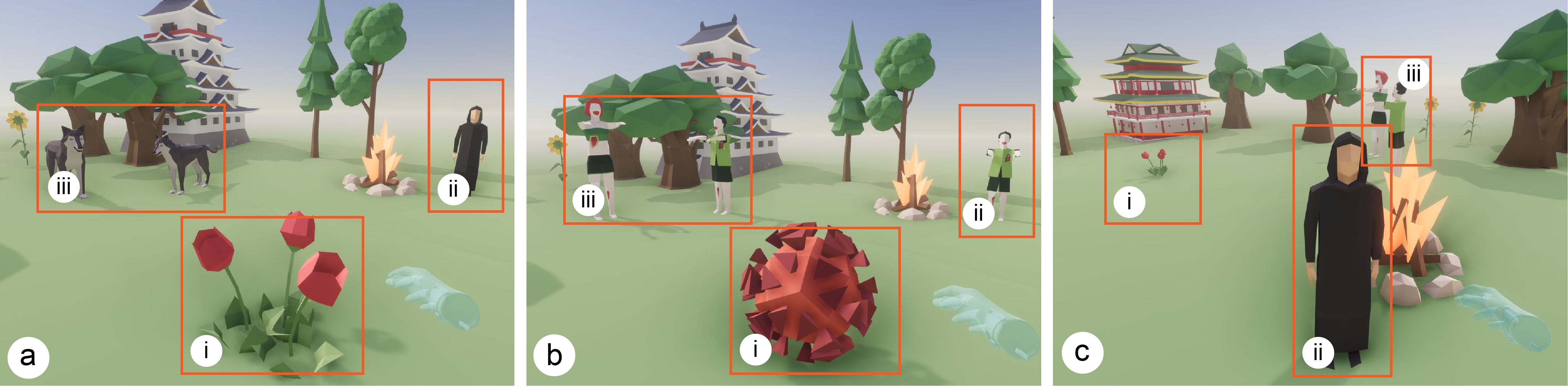}
\caption{An example interactive VR scene with different possible outcomes that depend on the player's interactions with the scene elements. The scene shown is composed using EntangleVR++'s visual interface. Left: (a) the initial scene state that a player sees before interacting with any elements. Three interaction options and resulting story pathways are available where a flower (i) turns into a virus when touched by the player; a man in black (ii) is saved from becoming a zombie if the player chooses to visit him before exposing the virus hidden in the flower; and a wolf pack (iii) is affected by the player's interactions with either the flower or the man in black, turning them into zombies. Middle: (b) the player chooses to touch the flower which unveils a hidden virus (i), causing the man in black (ii) and the wolf pack (iii) to turn into zombies immediately. Right: (c) the player chooses to interact with the man in black first which save this character from becoming a zombie (ii) but still turns the wolf pack into zombies (iii). In (c), the player interacts with the flower (i) and discovers the virus behind the zombies.}
\label{fig:teaser}
\end{figure}

Existing methods for creating interactive VR experiences, where the outcomes are based on the choices a player makes, require considerable time and effort. This is usually accomplished with professional game development engines such as Unity \footnote{https://unity.com/} or Unreal \footnote{https://www.unrealengine.com/en-US}. While specific VR authoring tools have recently been introduced \cite{3dpallete, wojciechowski2004, ruis, Dorner2014, vrforge2019}, they provide very limited support to add logical behaviors to the virtual elements in a scene \cite{Dorner2015}. Thus, creators may find themselves without many options to build complex VR experiences with branching narratives that are driven by cause-effect relationships with outcomes that depend on choices players make during gameplay. 

To simplify creating VR experiences where a player's choices related to interactions with virtual elements lead to different narrative outcomes (e.g., changing a series of object properties and states), we developed the interactive virtual scene composer, EntangleVR++. This tool is the next version of a previously presented system EntangleVR \cite{entanglevr}. Specifically, the newer version includes an improved interface (see Section~\ref{sec:system} for differences) to build interaction-driven experiences using a virtual scene composer that includes objects that are \textit{entangled} in relationships with one other. Figure \ref{fig:teaser} shows an example VR scene created with EntangleVR++ that supports different outcomes based on the choices a player makes.

We designed a visual programming interface to enable creators to build complex VR narratives and introduced \textit{entanglement} to allow the creation of sequences of correlated behaviors. Prior work has explored the use of visual programming as a valid interface for creating VR experiences \cite{schiavoni2017virtual}. The EntangleVR++ interface integrates fundamental object-oriented programming concepts with high-level representation of virtual objects as qubits.
Entanglement was introduced as a potential way to simplify narrative management for the creator. 

Specifically, \textit{entanglement} is used to describe inter-object relationships where object or qubit states are mutually dependent on each other, as an alternative to traditional if-else statements. Different from a classical cause-effect relationship where one object's state is usually a trigger that has an effect on another object's state, the states of two entangled objects are mutually correlated without a defined causal order. A single interaction event from the player is taken as a measurement in quantum computing that affects an entire entangled group and collapses the object states in different ways. For complex multi-entity scenarios, the cause-effect links between objects can become difficult to predict and track. A chained series of events and triggers is usually accomplished with multiple if-else statements in programming. In contrast, \textit{entanglement} enables us to construct non-separable behaviors of entangled objects, allowing the creators to easily follow and control the sequences of cause-effect behaviors through the visual interface.

In this work, we wanted to understand how well our idea of \textit{entanglement} enables creators to build interactive VR narratives involving cause-effect relationships. We asked 16 people to use EntangleVR++ in a 90-minute study. Based on interviews, analyses of screen recordings, and questionnaire feedback, we evaluated the system, particularly the idea of \textit{entanglement} and its capacity to support creation of correlated behaviors of virtual objects. We summarize the evaluation outcomes into four themes: (1) expressive visual interface, (2) creation with entanglement concept, (3) diversity in creative activities, and (4) learning support for basic quantum concepts. From the results, we see that our visual implementation of entanglement is an understandable and usable alternative to if-else statements for managing the creation of cause-effect relationships. Overall, we see the potential of using quantum inspired concepts for not only creating interactive VR scenes, but also interactive storytelling, puzzle game design, and creative arts. From the extracted themes and the study, we further derive four authoring strategies for designers interested in building visual interfaces for interactively creating virtual scenes that include relational objects i.e., objects connected with each other in cause-effect relationships, and multiple outcomes driven by player interaction choices made during gameplay.

\section{Related Work}

EntangleVR++ is inspired by prior work in interactive digital narrative in VR, visual programming for VR authoring, and quantum computing for casual and creative activities.

\subsection{Interactive Digital Narrative in Virtual Reality}
Interactive digital narrative originates mainly from Interactive Fiction (IF) and Role Playing Games (RPG). With the help of digital technologies, it has now evolved into several new formats such as video games (e.g., What Remains of Edith Finch\footnote{https://edithfinch.com/}, Firewatch\footnote{https://www.firewatchgame.com/}), web documentaries (e.g., Becoming Human \footnote{http://www.becominghuman.org}, Water's Journey \footnote{http://theevergladesstory.org/}, and interactive films (e.g., Black Mirror: Bandersnatch \footnote{https://www.netflix.com/title/80988062}, My One Demand \footnote{https://www.imdb.com/title/tt5376106/}. Different from non-interactive narrative, interactive digital narrative often offers flexibility and freedom for the viewer to select their desired branches to advance the story and determine, to some degree, how the story world evolves. This ability to control or participate in the plot construction can give the viewer a higher level of agency, and therefore a stronger sense of immersion \cite{murray1997}. VR offers the option for natural interactions and can thus be a powerful medium for interactive digital narrative. In recent years, an increasing number of VR narrative experiences have emerged though most are linearly designed where the player interactions do not affect the narrative's outcomes such as Wolves in the Walls \footnote{https://www.fable-studio.com/wolves-in-the-walls}, The Line \footnote{https://www.oculus.com/experiences/quest/2685959161497510/} and Battlescar \footnote{https://www.battlescarfilm.com/}. A few recent VR experiences offer the player freedom to choose their own paths and storylines such as Boba Yaga \cite{Cutler2020} and the Key VR \footnote{https://thekey-vr.com/}. EntangleVR++ aims to help creators produce such interactive narratives in VR easily and quickly, without requiring programming skills. Taking inspiration from quantum entanglement, EntangleVR++ enables creators to create VR scenes with multiple entangled objects that lead to different possible outcomes depending on the order in which the player interacts with them.

\subsection{Visual Programming for VR Authoring}
Visual programming systems (VPS) offer an opportunity to enable novice and casual users to create complex programs by using graphical elements and little training \cite{MYERS199097}. Because the variable states and data flows of a program are visualized, visual programming is often used in educational environments to help students understand relatively difficult concepts \cite{tamilselvam, PintoLlorente, Krishnamoorthy}. Visualization and immediate feedback in a VPS matches with the graphical interface concept of What-You-eXperience-Is-What-You-Get (WYXIWYG) \cite{WYXIWYG}. In an ideal VR authoring tool, the creator would immediately see and experience the output during the creation process. The simplicity of creation provided by the combination of visual programming and WYXIWYG has led to the design of a few VR authoring tools and platforms in recent years. For example, Dreams \footnote{https://www.playstation.com/en-us/games/dreams/} is an immersive creation system on the PlayStation gaming console that is widely used by artists  to create interactive VR mini-game worlds via its visual programming interface and scene composing tools. Scenior is a visual scripting system capable of generating VR training scenarios accompanied by a VR editor to interactively control and edit the generated training content \cite{scenior}. Ivy is a spatially situated visual programming tool that allows users to link IoT objects and insert logic constructs with visualized real-time data flow to compose mixed reality experiences \cite{Barrett2017}. FlowMatic is another recent system that provides a reactive visual programming and immersive authoring interface that allows the creator to compose interactive VR experiences \cite{flowmatic}. EntangleVR++ leverages the benefits of visual programming as well as WYXIWYG interfaces to allow creators to interactively preview their composed visual programs without compiling and building. It supports a fast test and iteration process where artists can interact to see how multiple virtual objects are entangled in a scene and how they may produce different outcomes upon different interactions.

\subsection{Quantum Computing for Casual and Creative Activities}
Many researchers in fields such as cryptography, physics, and machine learning are actively using quantum computing as a way to surpass the current computation limits of classical computers \cite{aaronson_2013}. As there is a limited number of available quantum computers, most of the existing research and education activities are performed on cloud-based quantum computers (e.g., IBM Quantum\footnote{https://quantum-computing.ibm.com/}, Xanadu\footnote{https://www.xanadu.ai/}, Ionq\footnote{https://ionq.com/}) or quantum simulators (e.g., Q\#\footnote{https://marketplace.visualstudio.com/items?itemName=quantum.DevKit64}, Qiskit\footnote{https://qiskit.org/}. The interfaces to construct a quantum computing program are usually determined by the service provider and they are designed for specific scientific tasks. For casual users who are interested in learning the basics of quantum computing or understanding quantum phenomena, there are very limited options available. For example, Quantum Flytrap \cite{quantumflytrap} and Entanglion \cite{entanglion} are educational games. Zable et al.'s \cite{zable2020} VR educational system can help visualize quantum computing concepts though it cannot be used as a tool for exploratory and creative activities. EntangleVR++ offers a visual programming interface that includes operational nodes borrowed from quantum computing such as qubits and quantum gates along with the idea of entanglement. Quantum computing elements are visualized in a beginner friendly style that can help creators with zero prior knowledge to build an interactive VR narrative experience while simultaneously learning about the basics of quantum computing.

\section{System Overview}\label{sec:system}
EntangleVR++ allows creators to design relationships between objects in a VR scene through a node-based graph interface. It is built upon EntangleVR's reactive visual programming system that works inside the Unity game engine as a plugin but does not require creators to have prior knowledge of Unity. \added{EntangleVR utilizes the xNode library \footnote{https://github.com/Siccity/xNode} to render a custom user interface.} The system incorporates various creation features such as object instantiation, object property settings, relationship definitions, and a tutorial panel. EntangleVR++ inherits all the features supported by the Unity scripting API such as visual effects, animation, spatial audio, and 3D rendering. Rapid prototyping using pre-composed game objects (Unity Prefabs) is also supported. Custom prefab assets, if added to the EntangleVR++ resource directory, will be loaded as part of the available object library for creators to add into virtual scenes. This enables creators to use preferred game assets (e.g., 3D models, audio-source objects, animated characters, particle effects) and add entangled relationships to them for creating narratives with cause-effect behaviors. Below we describe the terminology used in this work and specific features of EntangleVR necessary to clarify updates made in EntangleVR++.

\begin{figure*}[!t]
  \includegraphics[width=1.0\textwidth]{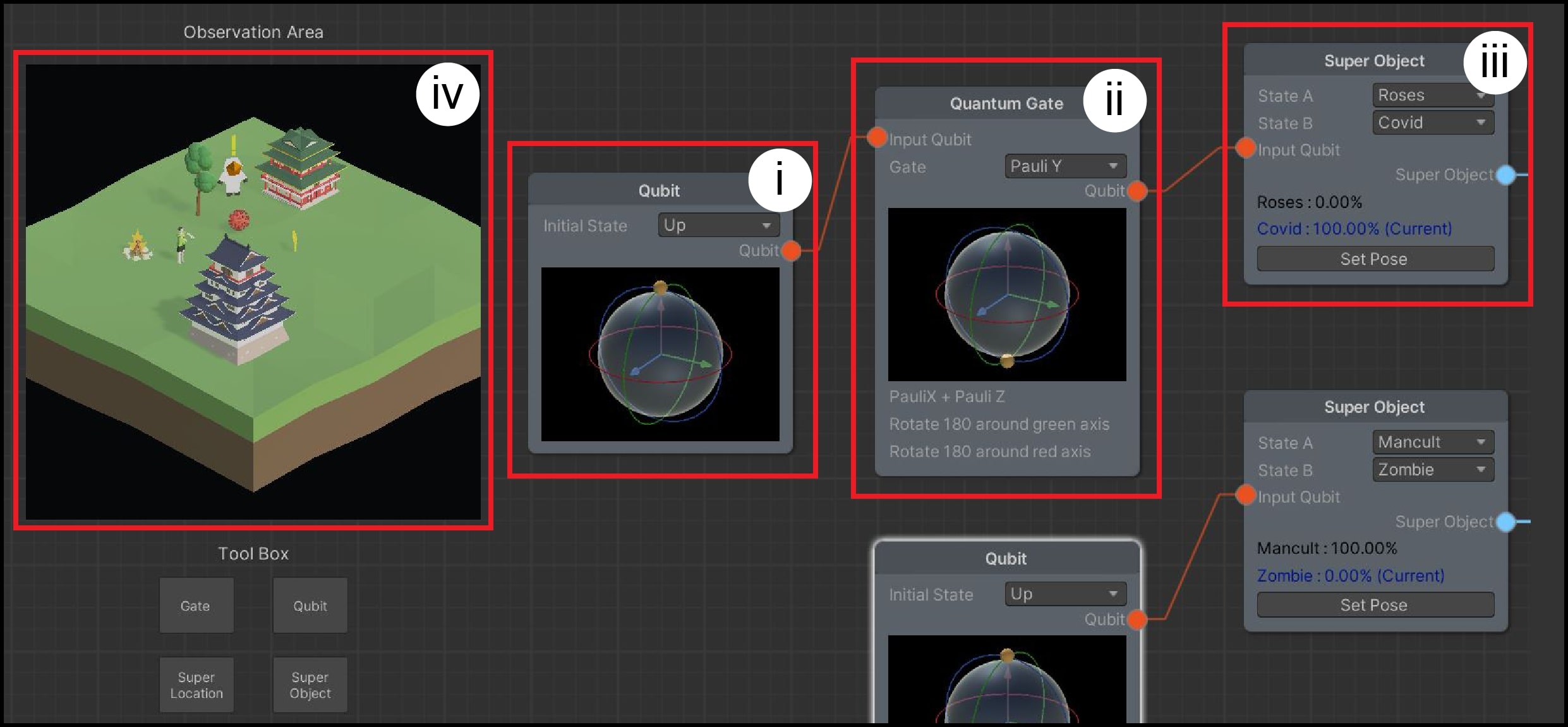}
  \caption{This figure shows EntangleVR++'s visual interface with quantum computing inspired nodes for interactive VR scene creation: (i) is a qubit node that lets the creator define initial probability amplitudes represented by a Bloch Sphere \cite{1946bloch}. (ii) is a quantum gate node that modifies the input qubit's state. (iii) is a super object node that take the input qubit value as a driving probability for its two possible object states. (iv) is an interactive preview window that shows a mini-map of all the virtual objects created in the scene in real-time and the creator can directly view and test the interaction outcomes by clicking on the appropriate interactable objects.}
    \label{fig:scenecreationflow}
\end{figure*}

\added{\subsection{Terminology}}
\added{In this work we use the following definitions of these terms:}
\begin{itemize}
    \item \added{\textbf{User / Creator} - the human artist or storyteller who desires to build interactive narratives in 3D (on screen) or VR using EntangleVR++.}
    
    \item \added{\textbf{Player} - the human who goes through the created narrative experience in 3D or VR.}
    \item \added{\textbf{Observation / Measurement} - the action that affects the properties of one or multiple virtual objects to change from quantum-like non-deterministic state to classical state. This action can be mapped to a player interaction (e.g., mouse click, VR controller touch) in a virtual environment.}
    \item \added{\textbf{Entanglement} - the non-separable shared state among multiple virtual objects, where the measurement results of each object (upon player interaction) consistently exhibit correlated relationships with the others.} 
\end{itemize}

\subsection{Interactive Scene Creation}
The scene creation process builds on EntangleVR \cite{entanglevr} that features a reactive visual programming interface to easily compose a virtual scene using quantum-inspired nodes. The nodes for scene creation include: Qubit, Quantum Gate, and Super Object (introduced in EntangleVR). In the original scene creation workflow, the creator starts with a qubit node as a basic unit of computation similar to how it is done in quantum computing. The creator can add as many quantum gates to the qubit as needed (e.g. Hadamard gate, Pauli X gate and Phase T gate) to modify the qubit into a desired state of specific probability amplitudes. After setting up the qubit, it can be used as an input to a Super Object node where the qubit's computational basis ($\ket{0}$ and $\ket{1}$) is mapped into two custom virtual object states defined by the creator. For example, a super object is defined by the creator to be in state A (a banana) and state B (an apple). The input qubit is set to have a 70\% probability to be measured at $\ket{0}$ and 30\% at $\ket{1}$. The representation of this super object in the virtual scene will therefore, have a 70\% probability to turn into a banana and 30\% probability to turn into an apple when a player tries to \textit{observe} this super object, i.e., interact with it by touching it during the VR experience. Figure \ref{fig:scenecreationflow} shows how the nodes are used and linked to create an interactive object that has a probabilistic state which is determined by the player's interaction. The creator can create as many super objects as they want in a VR scene.

\begin{figure}
  \includegraphics[width=\textwidth]{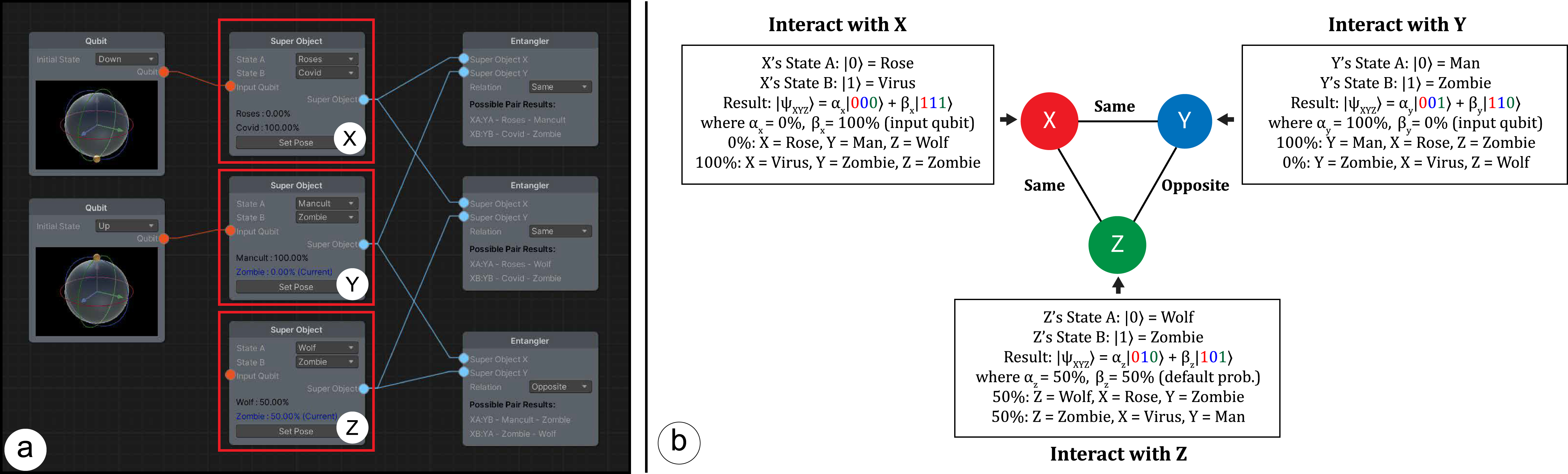}
  \caption{An example of how to build entangled relationships between three super objects in EntangleVR++. Left: (a) shows three different super objects (X), (Y), (Z) that are defined with individual input qubits and object states, and are linked with entanglers. Right: (b) shows a detailed explanation of how the graph in (a) enables the user to achieve different outcomes based on interactions with X, Y and Z. }
    \label{fig:multientangle}
\end{figure}

\subsection{Building Entangled Relationships}
Once the VR scene is populated by super objects, the creator can add Entangler nodes (introduced in EntangleVR) to build entangled relationships between the objects. An entangler node takes two super objects as input and creates a correlated relationship between their states when either one of the super objects is interacted with by the player. An \textit{observation} action is triggered on both super objects when a player approaches and interacts with any one of the entangled super objects, causing the two objects to simultaneously \textit{collapse} into outcomes based on the creator-defined relationship in the Entangler node. If the relationship is set to \textbf{Same}, the possible outcomes of an entangled pair will only be $\ket{00}$ or $\ket{11}$, and if it is set to \textbf{Opposite}, the outcomes will only be $\ket{01}$ or $\ket{10}$. 

A new addition to the Entangler node in EntangleVR++ is that multi-entity entanglement can be achieved by connections between three or more super objects to create more complex relationships that can increase the number of narrative branches. Depending on which super object a player first interacts with, the results of the \textit{collapse} from a multi-entity entanglement can vary considerably. The order of collapse is determined by the starting object following a breadth first traversal through all its connected nodes. Those objects that have already \textit{collapsed} into classical states will not \textit{collapse} again. This creates unique patterns of object state sequences based upon the player's interactions while providing the creator control over the outcomes. Figure \ref{fig:multientangle} shows an example of entanglement between three super objects and different outcomes from each interaction choice. 

\subsection{Creating Entangled Properties}
Another new addition in EntangleVR++ is that properties of super objects such as location and color can be defined in a probabilistic manner. Creators can link a super-property node (e.g., Super Location node, Super Color node) to the super object as an optional attribute. Similar to the two state definition on a super object node, a super property node allows the super object to have two properties with different probabilities that collapse when the super object is observed and interacted upon by the player. An input qubit can be provided to determine the probabilities of which property state the super object may get during its collapse. This added variety can allow for the creation of more dynamic interactive experiences.

\subsection{Real-time and Interactive Preview Workflow}
EntangleVR++ adds a new real-time preview window that immediately shows and reflects any changes in the visual program (add, remove, or edit nodes). An avatar can be added to the preview window by using a Character node to represent the player’s virtual body. Creators can choose between an on-screen preview avatar and a VR preview avatar. The on-screen preview spawns a mini character which can be controlled by mouse clicks like a third-person video game character. The VR preview avatar, on the other hand, drops a VR camera in the scene which provides a first-person view in the VR headset. Both avatars can be controlled by the creator to navigate around the created virtual scene while editing, to preview the outcomes of player interactions.

\section{Evaluation}
We conducted a user study with 16 remotely located participants over Zoom video conferencing software to evaluate how well EntangleVR++, and particularly the idea of \textit{entanglement}, supports the design of interactive VR narratives. We wanted to learn if \textit{entanglement} was easy to understand as implemented, and easy to use. Lastly, we wanted to get a sense of its potential to support creation of correlated behaviors of virtual objects to help guide the design of future versions of this design tool. 

As the goal of our evaluation was to gain insights into the interactive scene creation process using the concept of quantum entanglement, we did not require our participants to use a VR headset to experience the created virtual scenes. They were, instead, asked to complete scene creation tasks with the output of their creation visible in the interactive preview window. Before the study, we ran a pilot study with two remote participants to get early feedback on EntangleVR++'s functionality and user interface for identifying and fixing technical and user experience issues.

\subsection{Participants}
Sixteen participants (10 male, 6 female), aged between 18 - 44, were recruited through mailing lists of various academic departments on our campus. Since EntangleVR++ is designed for running on Windows, we looked for participants who had a PC desktop or laptop. Participants were from a wide range of backgrounds such as computer science (2), media arts (6), music (1), electrical engineering (6), and business (1), and they all had different levels of experience with programming and familiarity with quantum computing.

Participants' average rating of their knowledge of quantum physics or quantum computing was 2.8 (7-point Likert scale (1 = not at all, 7 = A great deal). 10 out of 16 participants had never taken any formal course on quantum computing in school or online. Participants also rated on average, 5.8, on a 7-point Likert scale (1 = not at all, 7 = A great deal) regarding their familiarity with scripting based programming. Python was the most common programming language reported (13 participants). The level of Python proficiency was not disclosed. On prior visual programming experience, participants rated on average of 3.0 on a 7-point Likert scale (1 = not at all, 7 = A great deal) with Max/MSP \footnote{https://cycling74.com/products/max} as the most common visual programming language reported (6 participants).

\subsection{Study Procedure}
\subsubsection{Pre-study}
Since EntangleVR++ is a plugin for Unity on Windows OS, participants were asked to install Unity game engine to ensure the VR authoring environment matched the development environment.

\subsubsection{Onboarding}
Each participant was onboarded in a Zoom video conference session. The total study session lasted 60 - 90 minutes, including the onboarding. At the start of the video call, we emailed the participant a consent form (study protocol approved by our Office of Research), a unique participant ID, a pre-study questionnaire, and a link to download the Unity package with the EntangleVR++ plugin. After providing informed consent, we guided participants to install EntangleVR++. After successful installation, participants were asked to complete two 15-minute long interactive learning sessions, each followed by a 10-minute scene creation task to evaluate the usability of the system for narrative creation and their understanding of entanglement for creating object relationships. \added{ The interactive learning sessions were self-paced and the participants were given the freedom to navigate through the tutorial step by step, with the option to revisit any section as needed for a deeper understanding.}

\subsubsection{Tutorials and Tasks}
The learning sessions had two parts: 1) learning how to create an interactive scene populated by super objects with super properties using qubits and gates (8 individual levels), 2) learning how to entangle multiple super objects and observing the different outcomes using the on-screen preview (4 individual levels). The first session had extra beginning levels that contained an overview of the visual programming interface. Participants also did short exercises at the end of each learning level to ensure familiarity and comfort with the interface, before proceeding to the study tasks.

\begin{table}[t]
\resizebox{\textwidth}{!}{%
\begin{tabular}{@{}llllll@{}}
\toprule

\multicolumn{5}{c}{\textbf{Task 1: Create super objects}} & \multicolumn{1}{c}{\textbf{Task 2: Create an entangled scenario}} \\ \midrule
\multicolumn{5}{l}{Step 1: Create two super objects}        & Step 1: Read the following scenario description:                  \\
\multicolumn{5}{l}{\begin{tabular}[c]{@{}l@{}}Step 2: Set your first super object with these properties: \\ a) 50\% chance of being a Horse, 50\% of being a Deer.  \\ b) Appears at coordinate (250, 250)\end{tabular}} &
  \begin{tabular}[c]{@{}l@{}}You are trying to get two items: an apple and a flower. \\ The apple is beneath a tree and the flower is right next to a spaceship. \\ However, if you try to touch the apple, a deer appears and destroys the flower. \\ When you try to get the flower, a zombie appears and eats the apple.\end{tabular} \\
\multicolumn{5}{l}{\begin{tabular}[c]{@{}l@{}}Step 3: Set your second super object with these properties:\\ a) 25\% chance of being a Zombie, 75\% chance of being a Spaceship \\ b) 86\% chance of appearing at coordinate (200, 200) \\ and 14\%  chance of appearing at coordinate (-320, -240)\end{tabular}} &
  \begin{tabular}[c]{@{}l@{}}Step 2: Connect the provided nodes and complete the graph using entangler.\\ Demonstrate the possible outcomes using the virtual character.\end{tabular} \\ \bottomrule
\end{tabular}%
}
\caption{Two tasks given to the participants in the user study.}
\label{tab:task}
\end{table}

A time-limited task was given at the end of each learning session. Table \ref{tab:task} shows the descriptions of the two tasks. We gave participants 10 minutes for each task and did not give them further instructions on how to complete the task. The task was based on what they had learned with a small challenge that required understanding of the implemented concepts to complete successfully. After the 10-minute limit, participants were allowed to ask for help if they were unable to complete the task. Participants were free to consult the learning instructions during the task or ask for any clarifications regarding the entanglement concept and the interface functions.

\subsubsection{Post-study}
After completing the two tasks, the participants were asked to fill out a post-study questionnaire about their experience learning and using EntangleVR++ to create a simple narrative experience. The questionnaire used a combination of standardized questions for usability \cite{Brooke1995} and custom questions to evaluate learnability and support for easy VR narrative creation on a 7-point Likert scale (1 = Strongly Disagree, 7 = Strongly Agree). We also conducted a one-on-one semi-structured interview at the end of the study that lasted about 15 minutes, to get more detailed feedback on the system and its use for creative purposes. 

\subsection{Data Collection}
Data was collected through screen recordings of their shared screen as participants followed the learning instructions and performed study related tasks. Additionally the semi-structured Zoom interview was recorded as were their responses to the pre- and post-study questionnaires. 

\begin{figure}
  \includegraphics[width=\textwidth]{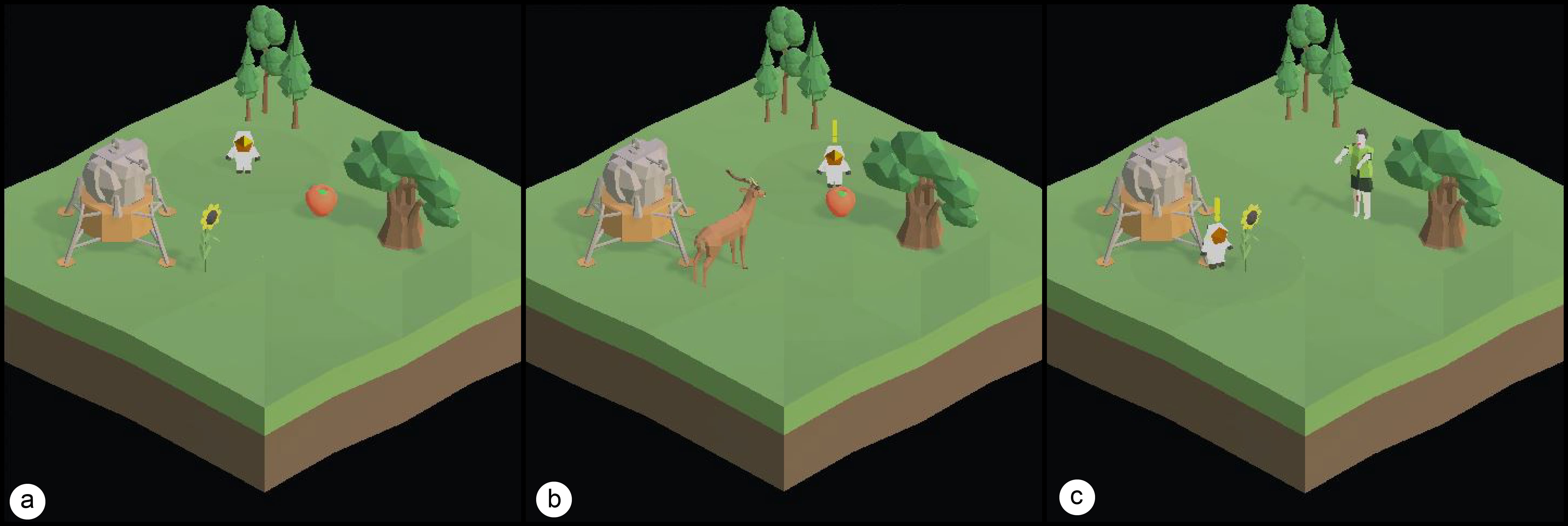}
  \caption{Visual results from Task 2 where participants were asked to create an interactive scene with two different outcomes based on player interaction. Left: (a) shows the initial scene state that has two target objects, an apple and a flower, as described in the task description given to the participants. Middle: (b) shows one outcome if the player chooses to interact with the apple that causes a deer to walk out from behind the trees and eat the flower. Right: (c) shows another outcome if the player chooses to pick up the flower first which causes a zombie to walk out from behind the lunar module and eat the apple. Outcomes (b) and (c) were created by the participants using entangled relationships between these objects (apple, flower, zombie, deer), resulting in a scenario where the apple and the flower cannot be taken by the player at the same time.}
    \label{fig:task2preview}
\end{figure}

\section{Results}
All participants successfully programmed Task 1 (qubit + super objects + gates). 13 participants completed programming Task 2 (qubit + super objects + entanglement) within the 10-minute time limit. The remaining three participants completed Task 2 after receiving a hint from the researcher. 

\subsection{Questionnaire Responses}
Post-study questionnaires are divided into three parts to separately evaluate: 1) usability 2) understanding of quantum entanglement concepts, and 3) support for easy creation of interactive VR scenes with entanglement concepts. 

\subsubsection{Usability}
We used a System Usability Scale (SUS) \cite{Brooke1995} to measure the the usability. Figure \ref{fig:SUS} shows separated results of positively worded items and negatively worded items based on the SUS scoring strategy \cite{Brooke2013}. 15 out of 16 participants agreed that EntangleVR++ was easy to use. All participants agreed that the various functions were well integrated. As we used a 7-point Likert scale for SUS \cite{finstad2010}, we converted the SUS score to a range of 0 - 100 with a 1.67 point multiplier. The overall SUS score with a $M = 80.83$ and $MAD = 9.66$ shows that our system has very good usability.

\setcounter{figure}{4}
\setcounter{subfigure}{0}
\begin{subfigure}
\setcounter{figure}{4}
\setcounter{subfigure}{0}
    \centering
    \begin{minipage}{0.45\linewidth}
        \includegraphics[width=\textwidth]{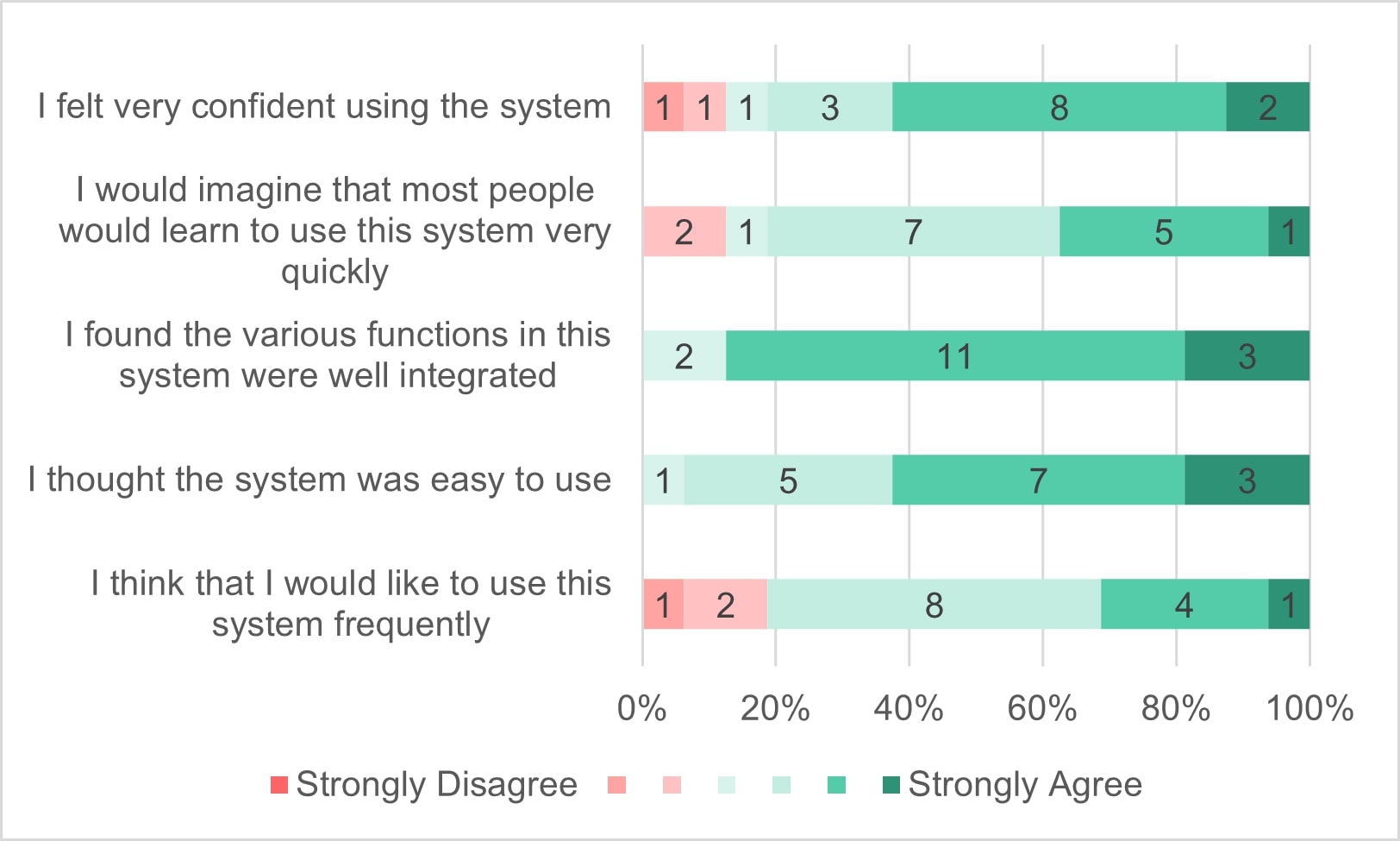}
        \centering
        \\(a)
        \label{fig:SUS1}
    \end{minipage}  
   \hspace{0.5cm}
\setcounter{figure}{4}
\setcounter{subfigure}{1}
    \begin{minipage}{0.45\linewidth}
        \includegraphics[width=\textwidth]{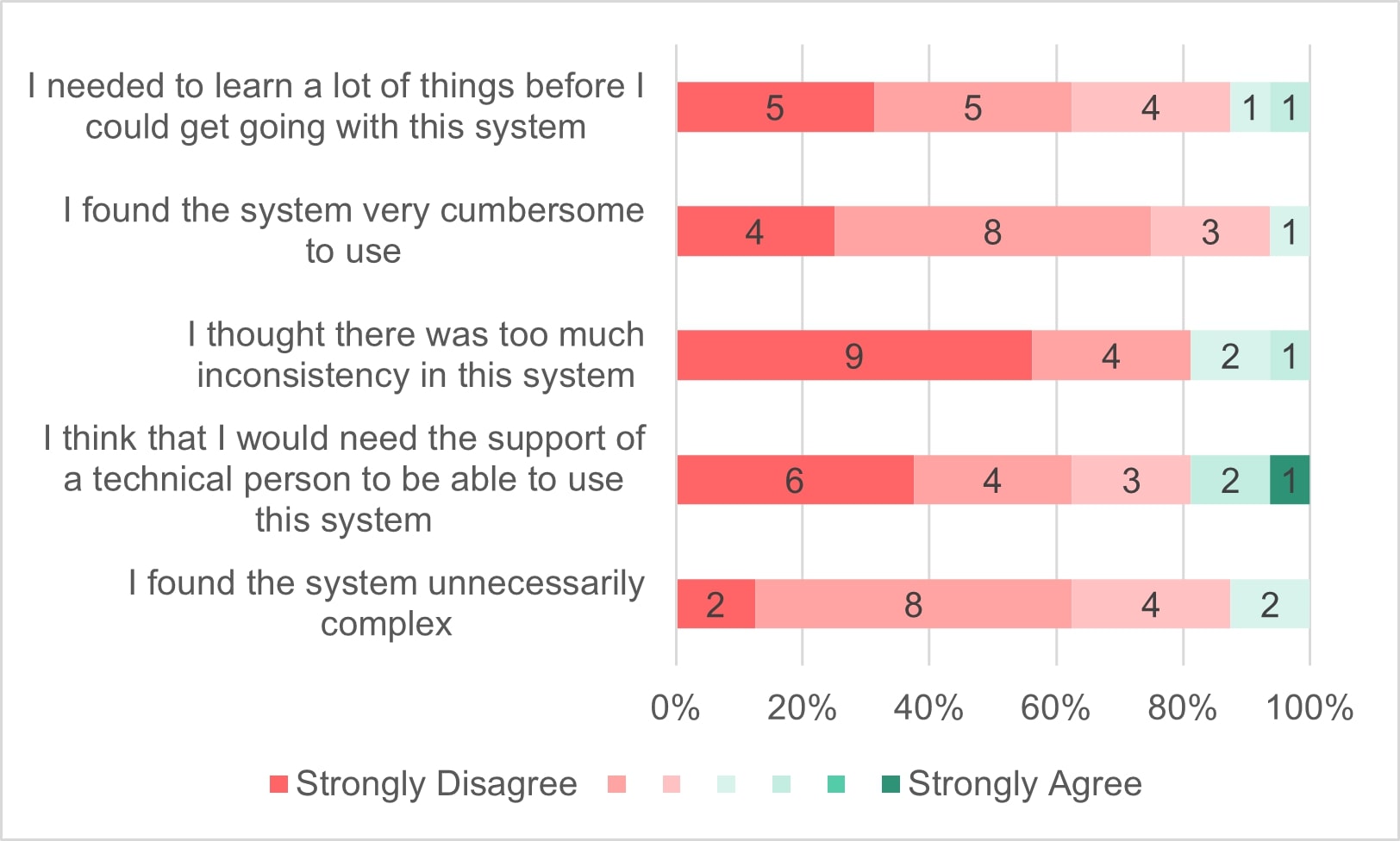}
        \centering
        \\(b)
        \label{fig:SUS2}
    \end{minipage}

\setcounter{figure}{4}
\setcounter{subfigure}{-1}
    \caption{Results of the System Usability Scale (SUS) questions. \textbf{(a)} shows questions 1, 3, 5, 7, and 9 (positively worded items). \textbf{(b)} shows questions 2, 4, 6, 8, 10 (negatively worded items)}
    \label{fig:SUS}
\end{subfigure}


\setcounter{figure}{5}
\setcounter{subfigure}{0}
\begin{subfigure}
\setcounter{figure}{5}
\setcounter{subfigure}{0}
    \centering
    \begin{minipage}{0.45\linewidth}
        \includegraphics[width=\textwidth]{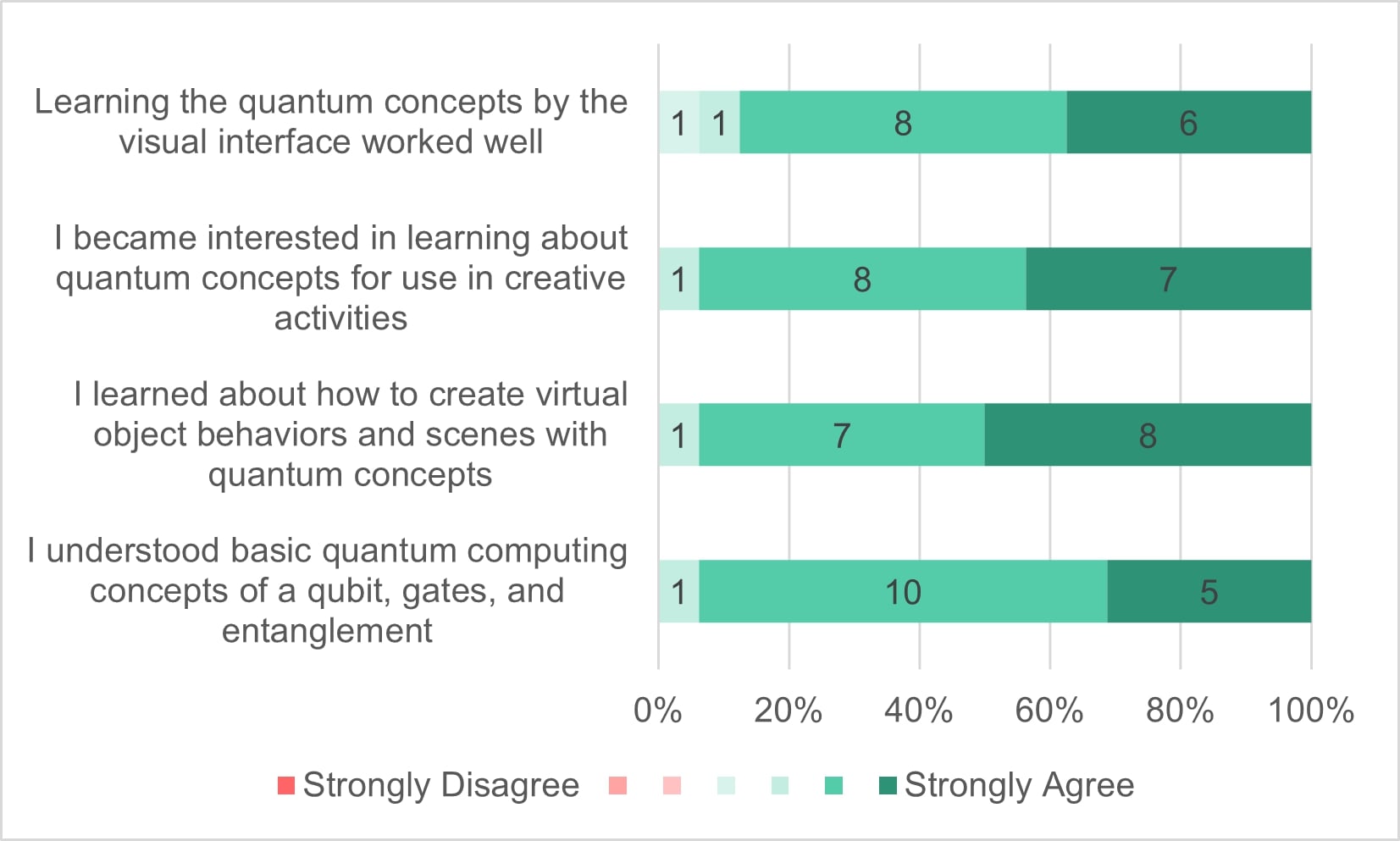}
        \centering
        \\(a)
        \label{fig:learnability}
    \end{minipage}  
   \hspace{0.5cm}
\setcounter{figure}{5}
\setcounter{subfigure}{1}
    \begin{minipage}{0.45\linewidth}
        \includegraphics[width=\textwidth]{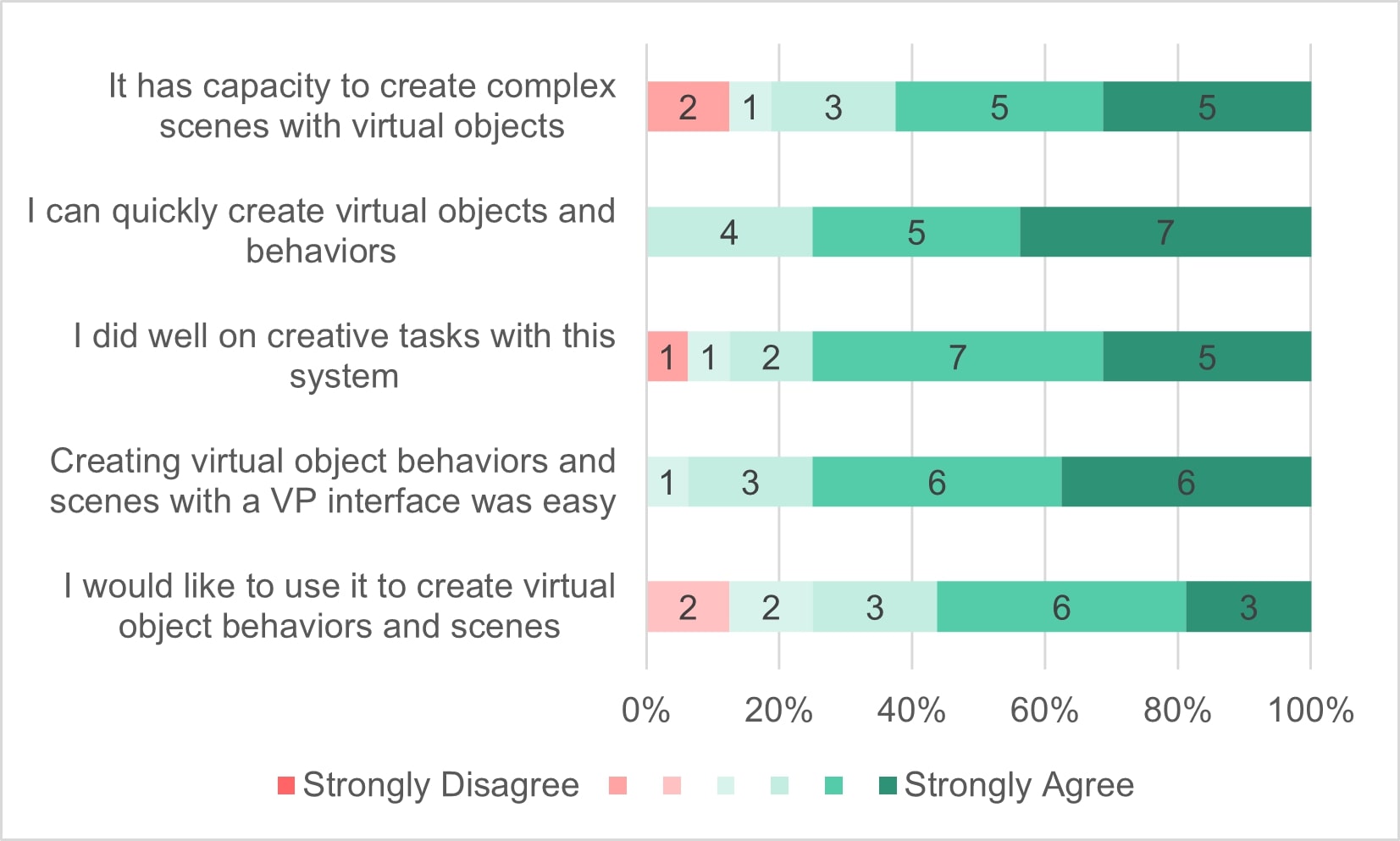}
        \centering
        \\(b)
        \label{fig:creativity}
    \end{minipage}

\setcounter{figure}{5}
\setcounter{subfigure}{-1}
    \caption{\textbf{(a)} Results from questions related to participant understanding of how quantum entanglement in used in the creation process. \textbf{(b)} Results from questions on system support for interactive VR scene creation}
    \label{fig:learncreate}
\end{subfigure}


\subsubsection{Understanding of quantum entanglement concepts}
Figure \ref{fig:learncreate} (a) shows the results from questions related to understanding of quantum entanglement concepts. All 16 participants reported feeling positive (average score of 6.31) about learning the quantum concepts and about their interest in using these concepts for creative activities. They also all agreed that the system successfully demonstrated its capacity to teach users how to create virtual object behaviors and scenes using quantum concepts. Of the 10 participants who had never taken any formal courses in quantum computing or quantum physics, all answered positively regarding their understanding of basic quantum concepts.

\subsubsection{Support for easy creation of interactive VR scenes with entanglement concept}
Figure \ref{fig:learncreate} (b) shows the results from questions related to interactive virtual scene creation. All the participants responded positively (average score of 5.79) to the system enabling quick creation of virtual objects and interactive behaviors. 13 participants said they believed the system can be used to create complex scenes and virtual objects. 15 thought that creating with a visual programming interface was easy. 12 participants expressed desire and willingness to use the system to create virtual object behaviors and scenes. 

\subsection{Post-study Interviews}
During the post-study interviews, we asked participants for feedback on the usefulness of the system, its weaknesses and their overall experience using it to create VR scenes, along with their thoughts on the application of entanglement for creativity. Interviews were transcribed by one researcher from the Zoom recordings and the text transcripts were exported for qualitative analysis. Screen recordings were matched with interview transcripts using the participants IDs. We employed an inductive thematic analysis approach \cite{braunclarke2006} to the interview texts. Two researchers independently reviewed the transcripts and derived their own codes and descriptions of the codes from the interview data. A discussion between the two researchers was held to further examine and refine the codes and jointly arrived at a set of four themes that are presented below. 

\subsubsection{Enhanced program understanding from an easy and expressive interface} \label{themes}
One of the most frequently mentioned positive features of EntangleVR++ was the ease of use. The reactive visual programming interface allowed outputs of different functions to be immediately visible in the preview window which helped the user understand the relationship between the super objects, entanglers and the corresponding outcomes. Ten participants mentioned that the expressive interface helped them easily understand the possible states of super objects and control the program design. They talked about how they ``enjoyed clicking and seeing the results right away''(P3) via the interactive preview for player interaction. Four participants reported that they did not face any difficulty during the entire learning and task completion process. At the same time, three participants pointed out that they are ``not a fan of visual programming'' (P0) or ``not very used to visual programming'' (P11) and would prefer a scripting interface, if available.

\begin{quotation}
P14: \textit{``it's very easy to use because of the functionality and usability the interface gives me. I think it is the nature of visual programming environment. Click, drag and drop directly renders on the scene. It is easy to use and understand what I'm doing, and see the effects of the actual output. This was quite great.''}
\end{quotation}

\subsubsection{Creation with entanglement and uncertainty}
Participants found the idea of associating interactive scene creation with a set of probabilistic events ``fresh and new''. More specifically, 50\% of the participants thought this method can be useful to ``simulate things that are very close to real life'' (P2) or to ``create a long drama of probabilistic chains of events or a choose-your-own-adventure story'' (P8) or ``a puzzle-like story'' (P11). The probabilistic nature of a qubit-driven super object in our system design is not just pure randomness, but something that is controllable with entanglement. P7 noted that ``because you can change the probabilities and then the entanglement can be used to create or build relationship between objects, I think that is very useful in [creating] visual content.''   

\begin{quotation}
P2: \textit{``The whole idea that allows to simulate virtual objects in quantum states is very interesting and figuring out how you can modify probabilities, let's say you can keep them random but still keeping their deterministic [behaviors], the concept of implementing these things that are close to real life was very fascinating.''}
\end{quotation}

\subsubsection{Diversity in creative activities} 
Participants pointed out a wide range of use cases like creating AR/VR narrative experiences, providing large variability in game character behavior and personality design, designing interactive media art installations where each player has a slightly different experience, using it to teach basic quantum concepts and to make music and soundscapes. Nine participants thought the system would be useful for making narrative based games. They talked about using entanglement for game design like ``creating dungeon-like loops with random but controllable sections'' (P9), ``choice-based games'' (P4), and ``behavior design of game character'' (P1). Three participants mentioned the system could be used by ``digital artists who have some programming experiences'' (P6) to create VR art installations. One participant explained the ability of the system to enable ``diverse various non-deterministic visuals'' as a reason for why artists would like using the system as it will ``reduce effort and time to build virtual environments and interactions between characters and objects'' (P14). One participant suggested using the system as a visual interface to an existing quantum computer. 

\subsubsection{Learning support for basic quantum concepts} 
Participants expressed that a system with expressive and interactive visual guides like ours can help people learn basic quantum concepts. While this is not the main goal of our prototype, users do need to understand the underlying concepts to be able to use the system to build interactive virtual scenes. Eight participants pointed out that students and beginner programmers unfamiliar with quantum concepts like qubit, quantum gate and entanglement could use the system in a learning environment (e.g. classroom). According to P6, the visual interface of the system was ``fun and interesting that it simplified quantum concepts to be really digestible and approachable.'' P5 described the system as ``a great way to introduce younger people and students in the classroom to dig into their creativity to make cool and compelling stuff.'' Participants who were familiar with quantum computing particularly found the visualization of concepts through nodes and the preview window to be a helpful refresher.

\begin{quotation}
P13: \textit{``Before today, I have minimal knowledge, but after using the system, I now know some of the concepts of quantum computing.''}
\end{quotation}

\section{Discussion: Authoring Strategies for Creators and Storytellers}
Our design demonstrates how concepts from quantum computing can inspire the design of a visual programming interface for interactively creating virtual scenes that include relational objects and multiple outcomes driven by player interactions. From the results, we see the potential of using quantum inspired concepts for creating interactive VR scenes. Visual programming provided an easy to use interface for engaging with complex interactive object relationships and we can imagine a creativity tool such as EntangleVR++ to potentially assist in the learning of difficult concepts.

We articulate four design strategies based on our interpretation of the data analysis and participant feedback and discuss our findings in this section. These strategies can serve as a guide for development and design of future quantum inspired authoring systems for creating interactive VR experiences. 

\subsubsection{Clear Visual Indication of Possible Interaction Outcomes}
One of the challenges in creating interactive narratives is that creators may feel intimidated when designing interactions that may lead to multiple possible outcomes. As the choice of interaction made by the audience is uncertain, after designing branching points that produce many different results, creators may start to lose their sense of control over how each interaction may contribute to the narrative outcome. For example, even simple binary choices made by a player in the Netflix show Black Mirror: Bandersnatch result in a trillion permutations of the narrative 
\footnote{\url{https://www.indiewire.com/2018/12/black\%2Dmirror\%2Dbandersnatch\%2Dendings\%2Done\%2Dtrillion\%2Dstory\%2Dcombinations\%2Dnetflix\%2Dstreaming\%2D1202031075/}}. Given this potential complexity for narrative creators emerging from simple binary choices, it is important to clearly keep them informed of the possible outcomes of each player interaction through the visual interface while simultaneously letting them test if the interaction produces the designed effects. EntangleVR++ displays the probabilities for each super object as well as the possible outcomes of other entangled super objects on each observer node so that the creators are fully aware of the impact of each interaction as they design them. We also provide a test button on the observer node to simulate an interaction trigger to let the creators see the actual outcome of a player interaction in the preview screen. 

\subsubsection{Provide Easy Walk-through for Testing Narrative Flow}
As an interactive narrative may have multiple possible paths or sequences of events that one can choose to go through, the creator may need to quickly test out these possible paths to see if they are consistent with the narrative flow. Although the creator may embed random or probabilistic elements in a story or embrace a more procedural style of narrative development, it may not be possible to test out all the cases. A quick and responsive way to let the creator easily walk through various major narrative paths is thus, highly desirable. In addition, game engines often provide play mode for the user to test the program, but it always starts at the very beginning of the experience. A customized starting point along with quick settings of object initial states can be helpful for the creator to run various test cases.

\subsubsection{Consider Techniques for Creating Sense of Plausibility}
Our participants were excited about using entanglement for the design of visual storytelling, interactive games, and digital art experiences. They were also able to propose extended usage of entanglement in interactive VR scenes and suggested more object properties such as animation, sound, scale, and character behavior that could be entanglement compatible. They thought that entangling and associating different object properties together for interaction effects could lead to a real-life like dynamism in an interactive experience. To us, this implied that the creators may want to provide the audience a sense of plausibility \cite{Slater2009PlaceIA} in their interactive narrative to increase the realism of the experience they are authoring. Therefore, a future authoring system should consider how to help the creator build an experience with plausible interactions and narratives. From our study, we can observe that uncertainty and probability-driven events are also considered useful strategies to simulate the unexpectedness of real life and may help increase the sense of plausibility. One participant specifically mentioned using a probabilistic distribution function to simulate things in real life, which can be considered the next step in creating plausible object behaviors. 

\subsubsection{Allow More Direct Manipulation}
The idea of direct manipulation has been explored in tool design for 2D visual art creation. Prior studies have shown that, for artists who are unfamiliar with programming concepts, manual manipulation and epistemic action on the target creation can help them better understand how the functions in the program behave \cite{hempelandchugh2016, jacobs2020}. Similarly, for programming virtual object behaviors, the creators may have a better understanding of their editing process if they can directly manipulate the object's interaction states while seeing a visual representation of perceivable changes. In our work, we used a Bloch sphere to visualize the probabilities of an super object with sliders to control its state, in addition to numeric value control of its state. During our study, we noticed that participants with lesser programming experience used the visual interface and the reactive outcome of slider control more than those with greater programming experience. They asked for even more interactivity and direct manipulation of super objects such as drag and drop features, to let them set more properties of objects in the scene. They also seemed to rely more on the interactive preview feature to test out their designs on the probabilities of super objects instead of trusting the numeric value representations, although it does not mean that they do not need the numeric reference. In contrast, accurate numeric control seemed to be more desired by those with greater programming experience, and they expressed explicit preference for a scripting interface over a visual programming interface. Not surprisingly, they tended to use the input field on the interface to type in an exact numeric value for object state control instead of using a slider to roughly reach the desired state. Therefore, it is useful to keep both ways of object state control to accommodate creators of different programming levels while permitting more direct manipulation.

\section{Limitations and Future Work}
Our work provides an updated implementation of the concept of quantum entanglement as a new method for creating interactive virtual scenes. However, our work has some limitations. First, we could only perform a short-term study using a guided tutorial followed by simple tasks for every participant. Participants did not have enough time to fully explore and create their own interactive narratives using the system. A long-term study enabling free form scene creation will be needed to see whether the study results still hold for both the system and the interface regarding support for easy creation of VR experiences. Second, future studies with two groups of users: with and without prior quantum knowledge, and with and without programming experience, may reveal newer aspects of learnability, creative support, and educational use of our system. \added{Third, a comprehensive exploration of prior experience in 3D content creation and engine-based development will enhance our understanding of how simplification of the quantum entanglement concept can be achieved, enabling us to effectively guide creators in utilizing our system.} Lastly, while EntangleVR++ has an expressive interface for interactive VR scene creation, our system can be expanded with more integration with the classical computing features, though beyond the scope of this paper, which could help make the system more powerful. 

Our main focus in this work was the exploration of the possibility of making interactive VR scene creation using the idea of quantum entanglement and to demonstrate that EntangleVR++ is capable of allowing beginner users to build relational virtual scenes that have multiple outcomes. In the future, we plan to continue exploring entanglement for different types of scene creation methods (e.g., procedural scene generation, multi-scene entanglement). \added{At the same time, we aim to implement and assess the effectiveness of our system through a workshop environment, providing a larger number of creators with the opportunity to learn and utilize EntangleVR++ to achieve their unique creative objectives. By evaluating real-world creative projects generated by our system, we intend to identify potential utility and user experience challenges inherent in the tool's design. This hands-on approach will enable us to gain valuable insights into the practical application of EntangleVR++ and refine its features based on user feedback.}

\section{Conclusion}
In this paper we presented an evaluation of EntangleVR++ to build interactive VR scenes that offer multiple different outcomes determined by choices an end user makes. To enable the connection between player interaction and scene narrative outcomes, we used the idea of quantum entanglement to manage inter-object relationships. We presented details of our system design and highlighted the new features we added to the original EntangleVR system. We conducted a user study with 16 participants to determine if the idea of quantum entanglement enabled the creation of understandable and manageable object relationships. Our results indicate that entanglement seems to work as an alternative to if-else statements and may make it easier for creators to manage complexity of branching narratives as the number of objects in a scene/experience increases. Participants were very positive and excited about using the concept of entanglement in their scene creation process and found EntangleVR++ to be a ``futuristic approach'' for VR scene creation.

\section*{Conflict of Interest Statement}
The authors declare that the research was conducted in the absence of any commercial or financial relationships that could be construed as a potential conflict of interest.

\section*{Author Contributions}
MC and MS developed the original idea. MC implemented the interface, executed the study, analyzed the data and wrote the manuscript. MP provided feedback and helped revise the manuscript. MS advised on the entire process, co-wrote the paper, and contributed to study design and data analysis. All authors approved the submitted version.

\section*{Funding}
This work was supported by the Human-AI Integration Lab at the University of California, Santa Barbara.

\section*{Acknowledgments}
We would like to thank Giuliana Barrios Dell'Olio, Bowen Zhang, and Yimeng Liu for providing feedback to the system interface and to the user study design.

\bibliographystyle{Frontiers-Harvard} 
\bibliography{reference}




\end{document}